\pdfoutput=1 
\documentclass[english,english,pra,english,preprint,amsmath,amssymb,aps,longbibliography,showkeys, titlepage]{revtex4-2}
\usepackage{lmodern}
\usepackage[T1]{fontenc}
\usepackage[latin9]{inputenc}
\usepackage{amsmath}
\usepackage{amsthm}
\usepackage{amssymb}
\usepackage{graphicx}

\makeatletter
\theoremstyle{plain}
\newtheorem{thm}{\protect\theoremname}
\theoremstyle{plain}
\newtheorem{prop}[thm]{\protect\propositionname}

\usepackage[T1]{fontenc}
\usepackage{amsmath}
\usepackage{amsthm}
\usepackage{amssymb}
\usepackage{graphicx}

\usepackage{graphicx}
\usepackage{float}
\usepackage{xcolor}
\usepackage[hypertexnames=false]{hyperref}
\hypersetup{
	breaklinks = true,
    colorlinks = true,
    citecolor = {blue},
	urlcolor = {blue},
	linkcolor = {blue}
}

\makeatother

\usepackage{babel}
\providecommand{\propositionname}{Proposition}
\providecommand{\theoremname}{Theorem}

\begin{document}
\title{Gromov ground state in phase space engineering for fusion energy}
\author{Hong Qin}
\email{hongqin@princeton.edu }

\affiliation{Princeton Plasma Physics Laboratory, Princeton University, Princeton,
NJ 08540}
\affiliation{Department of Astrophysical Sciences, Princeton University, Princeton,
NJ 08540}
\author{Elijah J. Kolmes}
\email{ekolmes@princeton.edu }

\affiliation{Princeton Plasma Physics Laboratory, Princeton University, Princeton,
NJ 08540}
\affiliation{Department of Astrophysical Sciences, Princeton University, Princeton,
NJ 08540}
\author{Michael Updike}
\email{michaelupdike@princeton.edu }

\affiliation{Princeton Plasma Physics Laboratory, Princeton University, Princeton,
NJ 08540}
\affiliation{Department of Astrophysical Sciences, Princeton University, Princeton,
NJ 08540}
\author{Nicholas Bohlsen}
\email{nb1738@princeton.edu }

\affiliation{Princeton Plasma Physics Laboratory, Princeton University, Princeton,
NJ 08540}
\affiliation{Department of Astrophysical Sciences, Princeton University, Princeton,
NJ 08540}
\author{Nathaniel J. Fisch}
\email{fisch@princeton.edu }

\affiliation{Princeton Plasma Physics Laboratory, Princeton University, Princeton,
NJ 08540}
\affiliation{Department of Astrophysical Sciences, Princeton University, Princeton,
NJ 08540}
\begin{abstract}
Phase space engineering by RF waves plays important roles in both
thermal D-T fusion and non-thermal advanced fuel fusion. But not all
phase space manipulation is allowed, certain fundamental limits exist.
In addition to Liouville's theorem, which requires the manipulation
to be volume-preserving, Gromov's non-squeezing theorem imposes another
constraint. The Gardner ground state is defined as the ground state
accessible by smooth volume-preserving maps. However, the extra Gromov
constraint should produce a higher-energy ground state. An example
of a Gardner ground state forbidden by Gromov's non-squeezing theorem
is given. The challenge question is: What is the Gromov ground state,
i.e., the lowest energy state accessible by smooth symplectic maps?
This is a difficult problem. As a simplification, we conjecture that
the linear Gromov ground state problem is solvable. 
\end{abstract}
\maketitle

\section{Phase space engineering for fusion energy and Gardner ground state}

Phase space engineering by RF waves plays important roles in both
thermal D-T fusion and non-thermal advanced fuel fusion, from plasma
heating \citep{Stix1992,Dodin2022} and current drive \citep{Fisch1987,Dodin2004,Fisch2003}
to instability suppression \citep{Reiman1983,Yoshioka1984,Haye2006,Reiman2018}
and $\alpha$-particle energy channeling \citep{Fisch1992,Fisch1995a,Fisch1995,Herrmann1997,Fisch2006,Ochs2015}.
Particle accelerator technologies frequently employ phase space engineering
to modify the characteristics of charged particles \citep{Courant1958,Chao93-all,Wangler98-all,Davidson01-all}.
Notably, methods exchanging the transverse and longitudinal emittance
of charged particle beams have been designed to enhance the beam quality
\citep{Wurtele99,Penn2000,Emma2006,Cornacchia2002,Ruan2011}.

Advanced fuel fusion using p-B11 or D-He3 needs to be operated in
a non-thermalized environment and thus requires significant power
circulation within the system to keep particles in non-equilibrium
energy states \citep{Nevins1998}. While this might seem inefficient,
it was recently illustrated \citep{Qin2024} that the energy doesn't
have to be lost if the power flow is carefully managed. This concept
is similar to energy recovery systems used in energy-recovering particle
accelerators~\citep{Schliessmann2023}. Just as successful deuterium-tritium
fusion requires near-perfect tritium recirculation with less than
0.1\% loss \citep{Clery2022,Abdou2020,Qin2024}, making advanced fuel
fusion work depends on maintaining non-thermal particle distributions
through efficient power recirculation in the system \citep{Qin2024}.

When using RF electromagnetic fields to manipulate charged particles,
certain fundamental limits exist. One of the constraints is imposed
by Liouville's theorem, which states that the volume particles occupy
in phase space must remain constant---you can reshape this volume,
but not compress it.

To highlight the importance of this and other constraints, we focus
on the energy extraction schemes for aneutronic fusion. While aneutronic
fusion has the advantage of releasing energy as charged particles
(which can theoretically be converted directly into electricity),
there's a catch. The initial fusing ions have much less energy than
the fusion products, meaning the released energy spreads out into
a larger phase space volume. This volume must be preserved during
any electromagnetic energy extraction process.

This raises a key question: For a given distribution of fusion products
(like alpha particles in proton-boron fusion), what's the maximum
energy we can extract electromagnetically? Put another way, what's
the lowest energy state we can reach through electromagnetic interactions?
Because of phase space volume conservation, this ``ground state''
energy can't be zero. As noted above, this defines a limit on the
energy that can be extracted from a plasma using RF waves. It also
has applications for understanding instabilities and turbulence, where
it quantifies the energy that is available to drive a mode \citep{Helander2017,Helander2020,Mackenbach2022,Mackenbach2023Measure,Mackenbach2023Miller,Kolmes2024Flutes,Kolmes2024,Helander2024}.

Gardner \citep{Gardner1963} posed the following problem: for a given
distribution of charged particles and an energy function, what is
the ground state (minimum energy state) accessible under volume-preserving
maps? He constructed the ground state by minimizing the system energy
under the constraint of constant phase space volume. This method became
known as the Gardner restacking algorithm \citep{Dodin2005,Kolmes2020}.
For any given two compact, connected sets that are diffeomorphic and
of the same volume in phase space, it can be proven (see the Appendix)
using a technique known as Moser's trick \citep{Moser1965} that there
must exist a volume-preserving diffeomorphism between the two sets.
Thus the ground state constructed by the Gardner restacking algorithm
is accessible by smooth volume-preserving maps.

However, volume preservation isn't the only constraint we need to
consider. A more stringent constraint is ``non-squeezability'',
which comes from the underlying symplectic nature of charged particle
dynamics in electromagnetic fields. We identify here that this means
that the Gardner ground state might not actually be achievable using
real electromagnetic fields, whether externally applied or self-generated
by the system. 

\section{Gromov's non-squeezing theorem}

For a canonical Hamiltonian system of $n$ degree of freedoms in $\mathbb{R}^{2n}$,
the dynamics is governed by the familiar Hamilton's equation 
\begin{align}
\dot{q}^{i} & =\frac{\partial H}{\partial p_{i}},\,\,\,\dot{p}_{i}=-\frac{\partial H}{\partial q^{i}},\label{eq:HE}\\
H & =H(q^{i},p_{i},t),\nonumber \\
i & =1,2,...,n,\nonumber 
\end{align}
The solution map of Eq.\,(\ref{eq:HE}) $\varphi_{t}:\boldsymbol{z}(0)=\left(q(0),\boldsymbol{p}(0)\right)\mapsto\boldsymbol{z}(t)=\left(\boldsymbol{q}(t),\boldsymbol{p}(t)\right)$
is symplectic, i.e., 
\begin{gather*}
\left(D\varphi_{t}\right)^{T}J\left(D\varphi_{t}\right)=J,\\
D\varphi_{t}\equiv\frac{\partial\varphi_{t}(\boldsymbol{z})}{\partial\boldsymbol{z}},\\
J=\left(\begin{array}{cc}
0 & I_{n\times n}\\
-I_{n\times n} & 0
\end{array}\right),
\end{gather*}
Here, $D\varphi_{t}$ is the Jacobian matrix of the solution map $\varphi_{t}$,
and $J$ defines an almost complex structure on $\mathbb{R}^{2n},$
i.e., $J:\mathbb{R}^{2n}\rightarrow\mathbb{R}^{2n}$ and $J^{2}=-1.$
Symplecticity is the defining characteristic of Hamiltonian systems,
and being symplectic is a much stronger geometric constraint than
being volume-preserving.

\begin{figure}[ht]
\centering \includegraphics[width=14cm]{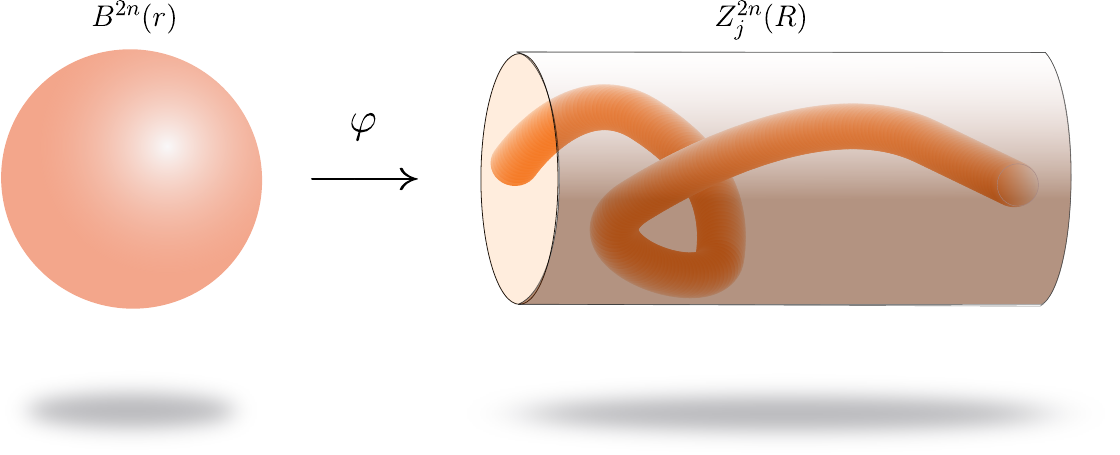} \caption{Gromov's non-squeezing theorem. No smooth symplectic map $\varphi$
in $\mathbb{R}^{2n}$ can send the ball $B^{2n}(r)$ to a cylinder
$Z_{j}^{2n}(R)$ when $r>R.$ But when $r=R,$ the ball $B^{2n}(R)$
fits comfortably within the bounds of the cylinder $Z_{j}^{2n}(R)$
without any squeezing.}
\label{B2Z}
\end{figure}

One way to characterize the symplectic constraint is given by Gromov's
non-squeezing theorem \citep{Gromov1985,Stewart1987,Hofer1994,deGosson06,Gosson2009,Gosson2009a,Gosson2011,Gosson2013,Gosson2016},
which states that there exists no smooth symplectic map $\varphi$
in $\mathbb{R}^{2n}$ sending the ball $B^{2n}(r)$ to a cylinder
$Z_{j}^{2n}(R)$ when $r>R,$ see Fig.\,\ref{B2Z}. Here, the ball
and the cylinder are defined as
\begin{align*}
B^{2n}(r) & \equiv\left\{ \left(q^{1},q^{2},...,q^{n},p_{1},p_{2},...,p_{n}\right)\left|\sum_{i=1}^{n}\left(p_{i}^{2}+q^{i2}\right)<r^{2}\right.\right\} ,\\
Z_{j}^{2n}(R) & \equiv\left\{ \left(q^{1},q^{2},...,q^{n},p_{1},p_{2},...,p_{n}\right)\left|p_{j}^{2}+q^{j2}<R^{2}\right.\right\} .
\end{align*}
The theorem significantly reduces the space of allowed manipulations
in phase space. Because $B^{2n}(R)\subset Z_{j}^{2n}(R),$ the ball
$B^{2n}(R)$ fits comfortably within the bounds of the cylinder $Z_{j}^{2n}(R)$
without any squeezing. However, if we attempt to expand the ball's
radius by even an infinitesimal increment, no symplectic map can force
this slightly larger ball into the same cylindrical space. This geometric
constraint presents a fundamental challenge---imagine a carpenter
trying to fit an ever-so-slightly enlarged piece into a predefined
space, only to find it mathematically impossible. 

With this constraint in mind, we pose the following problem of Gromov
ground state \citep{Qin2024}: for a given distribution of charged
particles and an energy function, what is the ground state under all
possible smooth symplectic maps? This Gromov ground state is higher
than the Gardner ground state. It determines the theoretical upper
limit of electromagnetically extractible energy \citep{Helander2017,Kolmes2020a,Helander2020,Kolmes2022}
in aneutronic fusion devices \citep{Bussard1994,Rostoker1993,Rostoker1997,Nevins1998,Mazzucato2023,Lerner2023,Kong2023,wei2023,LiuXie2024}.

Gromov's non-squeezing theorem suggests transitioning from volume-preserving
numerical algorithms \citep{Boris70,Stoltz02,Penn03,Qin2013Boris,he2015volume,zhang2015volume,He2016HigherRela,Tu2016}
to symplectic algorithms \citep{Qin2008VI-PRL,Squire4748,squire2012geometric,Xiao2013,xiao2015explicit,xiao2015variational,he2015Hamiltonian,he2016hamiltonian,zhang2016explicit,xiao2016explicit,qin2016canonical,kraus2017gempic,burby2017finite,Morrison2017,zhou2017explicit,xiao2017local,Xiao2018review,Xiao2019Relativisitic,Xiao2019field,Xiao2020Slow,Xiao2021Explicit,Glasser2020,Wang2021,Kormann2021,Perse2021,Glasser2022,CamposPinto2022,Burby2023}
to more accurately simulate the phase space engineering processes
of charged particles.

\section{Gardner ground state forbidden by Gromov's non-squeezing theorem\label{sec:GG}}

In this section, we demonstrate an example of a Gardner ground state
that is not a Gromov ground state.

Consider a system with 2 degrees of freedom in $\mathbb{R}^{4}$ with
canonical coordinates $(x,y,v_{x},v_{y}).$ Assume there is an external
potential in the $x$-direction, 
\[
\phi(x)=\frac{1}{2}x^{2}.
\]
A particle's energy is 
\[
\varepsilon=\frac{1}{2}\left(x^{2}+v_{x}^{2}+v_{y}^{2}\right).
\]
This system is a simplified model for charged particles dynamics in
accelerators or quadrupole ion traps \citep{Davidson01-all}.

Suppose the initial distribution function $f_{0}$ is uniform inside
$B^{4}(r),$ i.e., 
\begin{align*}
f_{0}\left(x,y,v_{x},v_{y}\right) & =\Theta\left(r^{2}-\left(x^{2}+y^{2}+v_{x}^{2}+v_{y}^{2}\right)\right)\\
 & \equiv\begin{cases}
1, & \left(x^{2}+y^{2}+v_{x}^{2}+v_{y}^{2}\right)<r^{2},\\
0, & \textrm{otherwise}.
\end{cases}
\end{align*}
Here, $\Theta(x)$ is the Heaviside step function. Such a distribution
function is known as a water-bag distribution. The phase space volume
occupied by $f_{0}$ is 
\[
V(f_{0})=\int_{B^{4}(r)}dxdydv_{x}dv_{y}=\frac{\pi^{2}}{2}r^{4}.
\]
The energy of the system is 
\begin{align*}
W(f_{0}) & =\int\frac{1}{2}\left(x^{2}+v_{x}^{2}+v_{y}^{2}\right)f_{0}dxdydv_{x}dv_{y}.\\
 & =\int_{B^{4}(r)}\frac{1}{2}\left(x^{2}+v_{x}^{2}+v_{y}^{2}\right)dxdydv_{x}dv_{y}\\
 & =2\int_{0}^{\pi/2}W_{3}(r\cos\theta)r\cos\theta d\theta\\
 & =2\int_{0}^{\pi/2}\frac{2\pi}{5}r^{6}\cos^{6}\theta d\theta=\frac{\pi^{2}}{8}r^{6},
\end{align*}
where 
\[
W_{3}(r)\equiv\int_{0}^{r}\frac{a^{2}}{2}4\pi a^{2}da=\frac{2\pi}{5}r^{5}
\]
is the energy line density inside the ball $B^{3}(r)\equiv\left\{ \left(x,v_{x},v_{y}\right)\left|x^{2}+v_{x}^{2}+v_{y}^{2}<r^{2}\right.\right\} $
.

The energy of the system can be reduced by pushing particles to regions
in phase space with lower energy density while preserving the phase
space volume, or the volume of the water bag. In particular, we can
squeeze, via a volume-preserving map, the initial ball $B^{4}(r)$
into the following ``cylinder'' in $\mathbb{R}^{4}$, 
\[
B^{3}(R)\times L=\left\{ \left(x,y,v_{x},v_{y}\right)\left|x^{2}+v_{x}^{2}+v_{y}^{2}<R^{2}\text{ and \ensuremath{0<y<L}}\right.\right\} .
\]
Our intention is to mold the ``cylinder'' into a long noodle along
the $y$-axis by increasing $L$ and decreasing $R$ while maintaining
the same phase space volume.

The distribution function is now 
\begin{align*}
f\left(x,y,v_{x},v_{y}\right) & =\begin{cases}
1, & \left(x,y,v_{x},v_{y}\right)\in B^{3}(R)\times L,\\
0, & \textrm{otherwise}.
\end{cases}
\end{align*}
The phase space volume of the system is 
\[
V(f)=\int_{B^{3}(R)\times L}dxdydv_{x}dv_{y}=\frac{4\pi}{3}R^{3}L.
\]
The energy of the system is 
\begin{align*}
W(f) & =\int\frac{1}{2}\left(x^{2}+v_{x}^{2}+v_{y}^{2}\right)fdxdydv_{x}dv_{y}.\\
 & =\int_{B^{3}(R)\times L}\frac{1}{2}\left(x^{2}+v_{x}^{2}+v_{y}^{2}\right)dxdydv_{x}dv_{y}\\
 & =W_{3}(R)L=\frac{2\pi}{5}R^{5}L=\frac{3}{10}R^{2}V.
\end{align*}
Volume-preserving requires $V$ is a constant, i.e., 
\[
V=\frac{\pi^{2}}{2}r^{4}=\frac{4\pi}{3}R^{3}L=\text{const.}
\]
Under this constraint of constant volume, we can mold the ``cylinder''
into a long noodle by letting $R\rightarrow0$ and $L\rightarrow\frac{3V}{4\pi R^{3}}$,
which leads to 
\[
W(f)\rightarrow0.
\]
Thus, the ground state energy is $0,$ which is reachable when $R\rightarrow0$,
if volume-preserving is the only constraint. Let's call this ground
state Gardner ground state.

However, this Gardner ground state is not reachable by symplectic
maps, because this volume-preserving map sends $B^{4}(r)$ to 
\[
B^{3}(R)\times L\subset Z_{1}^{4}(R)\equiv\left\{ \left(x,y,v_{x},v_{y}\right)\left|x^{2}+v_{x}^{2}<R^{2}\right.\right\} .
\]
According to Gromov's non-squeezing theorem, this map can't be symplectic
if $R<r.$

\section{What is the Gromov ground state?}

The unanswered question is: What is the ground state accessible by
smooth symplectic maps? Let's call such a ground state the Gromov
ground state. Can the energy of the Gromov ground state be close to
the energy of the Gardner ground state?

Assume there is a symplectic map sending $B^{4}(r)$ to $B^{3}(R)\times L$
(this assumption is very likely to be wrong). Then the minimum $R$
allowed by Gromov's theorem for symplectic maps is $r.$ If we take
this $B^{3}(r)\times L$ with $L=3V/4\pi r^{3}=3\pi r/8$ to be an
``approximate Gromov ground state'', then its energy would be 
\[
W(f)=\frac{3}{10}r^{2}V=\frac{3\pi^{2}}{20}r^{6}=\frac{6}{5}W(f_{0}).
\]
Obviously, this ``approximate Gromov ground state'' is not a good
approximation at all, because its energy is not reduced from the initial
energy of $f_{0}$. This demonstrates that Gromov's non-squeezing
theorem puts a strong constraint on the ground state accessible by
symplectic maps relative to the volume-preserving maps. The noodle
allowed by Gromov's theorem looks more like a ball because $L=3\pi r/8\sim R=r.$
That is why the energy of the state is not reduced relative to the
initial ball. In other words, the volume-preserving constraint allows
the ball to be molded into a long noodle. But Gromov's theorem says
the footprint in the $(x,v_{x})$ plane and the $(y,v_{y})$ plane
can't be reduced, and if one must mold the ball into a noodle, then
the noodle has to be thick and short.

On the other hand, Gromov's theorem does not prohibit molding the
ball into a mushroom with a long stem and a thin cap. This is because
this mushroom's footprint in the $(x,v_{x})$ plane and the $(y,v_{y})$
is not reduced relative to the initial ball. This mushroom's energy
could be very close to $0$, since the cap can be very thin. However,
the issue is that Gromov's theorem only says squeezing is not allowed,
it does not say what is accessible via symplectic maps. This mushroom
may still be inaccessible via symplectic maps, even though it satisfies
Gromov's non-squeezing constraint.

Another argument against this mushroom being a Gromov ground state
is the non-squeezing theorem applied to the stem by itself. The pre-image
of the stem should be almost the entire ball, which can't be squeezed
into the stem. In this example, the pre-image of the stem is a neighborhood
of the ball.

It was proven \citep{Katok1973} that a smooth symplectic map exists
to send a set in phase space to an arbitrarily close neighborhood
of another set of the same phase space volume if the derivatives of
maps are allowed to be arbitrarily large. Therefore, the Gromov ground
state approaches the Gardner ground state if the symplectic maps are
allowed to have arbitrarily large derivatives. Of course, arbitrarily
large derivatives implies that the map is becoming non-smooth.

Interestingly, this suggests connections with the theory of diffusively
accessible free energy, in which Gardner's restacking operation (which
exchanges the populations of two elements of phase space) is replaced
with a mixing operation (wherein their populations are averaged) \citep{Fisch1993,Hay2015,Hay2017,Kolmes2020a,Kolmes2020,Kolmes2022,Kolmes2024}.
Suppose there existed a symplectic map that produced very fine-scale
structure in some local region in phase space, such that the map appeared
to produce diffusion when viewed on a coarser scale in phase space
(this is essentially the intuition behind processes like quasilinear
diffusion). Suppose, furthermore, that such a map could be applied
to different, perhaps overlapping regions of phase space, again with
the effect of generating fine-scale structures that appear to produce
diffusion on larger scales. Then, in the limit where these fine-scale
structures could be made arbitrarily fine, it would be possible to
construct a symplectic map to a state arbitrarily close to any state
that is accessible through mixing operations. It has been shown \citep{Kolmes2020}
that sequences of mixing operations can access states that are arbitrarily
close to the Gardner ground state, so this would imply that the free
energy accessible through symplectomorphisms is arbitrarily close
to the free energy accessible through volume-preserving maps. The
interesting point here is that the requirement of arbitrarily fine-scale
structure suggests that such a symplectomorphism would have very large
derivatives, consistent with the result in Ref~\citep{Katok1973}.

In any event, the result of Ref.~\citep{Katok1973} suggests that
the free energy accessible through an arbitrary symplectomorphism
is arbitrarily close to that accessible through volume-preserving
maps, but that the symplectomorphisms needed to accomplish this may
not be smooth and therefore may not be appropriate in all scenarios.
With this consideration, we should specify the classes of allowed
symplectic maps when posing the Gromov ground state problem. Since
real fusion devices are of finite size, we could also further require
the domain and range of the symplectomorphisms to be bounded.

As an example of practical importance, we can study the linear Gromov
ground state problem. Most, if not all, beam optical components for
controlling charged particles can be modeled by linear symplectic
maps \citep{Courant1958,Chao93-all,Wangler98-all,Davidson01-all},
even though linear symplectic maps do not describe the wave-induced
dynamics for current drive \citep{Fisch1987} and $\alpha$-channeling
\citep{Fisch1992,Fisch1995a,Fisch1995,Herrmann1997,Fisch2006,Ochs2015}
in tokamaks. It is a known result \citep{Gosson2016} that linear
symplectic maps also define a valid symplectic capacity, and it agrees
with the symplectic capacity defined by general symplectic maps for
balls and cylinders. Intuitively, this can be interpreted as that
linear symplectic maps are as flexible as nonlinear symplectic maps.
If so, a neighborhood of the Gromov ground state might be approachable
by linear symplectic maps. Let's call the minimum energy state accessible
via linear symplectic maps the linear Gromov ground state.

In particular, for the counter example given above, we pose the following
problem: for the external potential $\phi(x)$ and $f_{0}$ defined
in Sec.\,\ref{sec:GG}, what is the linear Gromov ground state? i.e.,
what is the $4\times4$ symplectic matrix $S$ that minimize the energy
of $S\left(B^{4}(1)\right)$? We conjecture that this problem is solvable
\citep{Updike2024}.

\section{Summary and Discussion}

What we have identified here is an important constraint on the ground
state energy of collisionless rearrangement by waves of charged particles
in a plasma. Indeed, this constraint applies to any rearrangement
by means of Hamiltonian dynamics. In particular, we suggest that applying
the nonsqueezing theorem of Gromov could lead to a higher ground state
energy, at least in practical cases. The point here that has gone
unrecognized in the existing literature on plasma available energy
is that Hamiltonian dynamics is always phase-space-volume-preserving,
but that not all phase-space-volume-preserving transformations are
accessible via Hamiltonian dynamics (or more formally, symplectic
maps).

Because of the interest in fusion applications of recovering particle
energy (particularly fusion byproduct energy) in waves, it is of great
interest to know the maximum recoverable amount or the ground state
for a given energy distribution. The first question is what the allowable
ground states are according to the rules of particle motion. For particles
obeying Hamiltonian dynamics, the rules are clear for getting from
configuration $A$ to configuration $B$ in the 6D phase space: One,
there must be phase space conservation, so configuration $B$ has
to have the same phase space densities as configuration $A$. This
leads to Gardner restacking \citep{Gardner1963,Dodin2005,Kolmes2020}.
But configuration $A$ has also to travel an allowable path to configuration
$B$. That leads to the Gromov non-squeezing constraint \citep{Dodin2005,Kolmes2020},
which is much harder to quantify. However, in view of the last section,
if transformations with arbitrarily large gradients are allowed, then
the Gromov ground state approaches the Gardner ground state.

However, not all allowable transformations are practical. In controlling
charged particles by waves in the most important fusion applications,
it is invariably the case that the waves are arranged to diffuse particles
in velocity space (like in current drive) or in the combined velocity-configuration
space (like in alpha-channeling). Because wave-particle interactions
are a blunt instrument, it is not practically possible to exert extremely
fine control over the particle rearrangement. This practical limit
is important; in the absence of this limit we might be allowed transformations
that are not smooth and therefore allow energy recovery approaching
the Gardner free energy, even taking into account the Gromov constraint
as indicated in the last section. But with only blunt transformations
possible, both the Gardner ground state energy and the Gromov ground
state energy must necessarily rise. However, the extent of this rise
may not be the same.

This same notion of ``bluntness'' -- in which the realistically
accessible states are constrained by a lack of perfect control over
which phase space volumes go where -- is closely related to the idea
of the free energy under diffusive operations \citep{Fisch1993}.
Interestingly, even for that problem, arbitrarily fine control over
the regions of phase space undergoing diffusion makes it possible
to replicate the Gardner ground state arbitrarily closely \citep{Kolmes2022}.

If arbitrary symplectomorphisms can replicate the Gardner ground state
arbitrarily well, that still leaves open the problem of finding the
accessible energy under the Gromov constraint when infinitely fine-grained
control over the map is not possible. In the absence of arbitrarily
fine control by the waves, Gardner restacking is also modified; one
could call the resulting free energy the \textit{coarse-grained Gardner
ground state energy} (one could imagine, for example, the Gardner
problem in which phase space is discretized with some finite cell
size). Of course, that will depend on how coarsely the plasma is put
into bins of constant density in the 6D configuration space. Now adding
the Gromov constraint on allowable particle motion, one can define
for such a degree of control by waves a \textit{coarse grained Gromov
ground state energy}. The \textit{coarse-grained Gromov ground state
energy} will necessarily be higher than the \textit{coarse-grained
Gardner ground state energy}. However, this is the energy that is
hard to compute, except in the case when infinitely fine-grained discretization
is allowed, in which case both energies approach the same ground state
Gardner energy.

But the practical question, at least for diffusion by waves for extracting
energy, is in fact what is the \textit{coarse-grained Gromov ground
state energy}. This energy not only depends on the coarseness, but
also on how coarseness is defined. One approach to this problem in
which we conjecture a solution may be approachable is by examining
the ground state under linear symplectic mappings, which should be
solvable.

\appendix* 

\section{Proof of accessibility of Gardner ground state via volume-preserving
map}

In this Appendix, we show that the Gardner ground state, constructed
by minimizing the system energy under the constraint of constant phase
space volume, is accessible by smooth volume-preserving maps. It suffices
to prove that given any two physically well-defined regions of the
same volume in phase space, we can always find a volume-preserving
diffeomorphism connecting the two regions. We formulate this result
as the following proposition.
\begin{prop}
\label{prop:vp}Let $A$ and $B$ are two compact, connected sets
of $\mathbb{R}^{m}.$ If $A$ and $B$ are diffeomorphic and have
the same volume as measured by a volume form $\Omega$ in $\mathbb{R}^{m},$
then there exists a volume-preserving diffeomorphism $\psi:A\rightarrow B$. 
\end{prop}

\begin{proof}
Let $\phi:A\rightarrow B$ be the diffeomorphism between $A$ and
$B.$ Since $\phi$ is not necessarily volume-preserving, $\left.\phi^{*}\Omega\right|_{A}\neq\left.\Omega\right|_{A}$
in general. However, a volume-preserving diffeomorphism $\psi:A\rightarrow B$,
defined by the property $\left.\psi^{*}\Omega\right|_{A}=\left.\Omega\right|_{A}$,
can be constructed as follows.

Let $\left.\Omega_{1}\right|_{A}=\left.\phi^{*}\Omega\right|_{A}$.
Because $A$ and $B$ have the same volume as measured by a volume
form $\Omega$, 
\[
\int_{A}\Omega=\int_{B}\Omega,
\]
which implies 
\[
\int_{A}\Omega=\int_{\phi(A)}\Omega=\int_{A}\phi^{*}\Omega=\int_{A}\Omega_{1}.
\]
According to Theorem \ref{thm:Moser}, on the compact, connected manifold
$A,$ there exists a diffeomorphism $\tau:A\rightarrow A$ such that
$\Omega=\tau^{*}\Omega_{1}.$ Let 
\[
\psi=\phi\circ\tau:A\rightarrow B.
\]
We have 
\[
\left.\psi^{*}\Omega\right|_{A}=\tau^{*}\circ\phi^{*}\Omega=\tau^{*}\Omega_{1}=\Omega.
\]
Thus, $\psi:A\rightarrow B$ is a volume preserving diffeomorphism
between $A$ and $B.$
\end{proof}
Here, we require the two sets to be diffeomorphic in addition to have
the same phase space volume. This is to rule out situations where
the two sets are topologically different. For phase space engineering
in the present context, the phase space $\left\{ \left(q^{1},q^{2},...,q^{n},p_{1},p_{2},...,p_{n}\right)\right\} $
is identified with $\mathbb{R}^{m}$ $(m=2n)$, and the volume form
is the canonical volume form $\Omega=dp_{1}\land...\wedge dp_{n}\wedge dq^{1}\wedge...\wedge dq^{n}.$
The proof of Proposition \ref{prop:vp} used the following theorem. 
\begin{thm}[Moser]
\label{thm:Moser}Let $\Omega$ and $\Omega_{1}$ are two volume-forms
on a compact, connected manifold $M.$ There exists a diffeomorphism
$\tau:M\rightarrow M$ such that $\Omega=\tau^{*}\Omega_{1}$ iff
$\int_{M}\Omega=\int_{M}\Omega_{1}.$ 
\end{thm}

Moser \citep{Moser1965} proved Theorem \ref{thm:Moser} using a technique
that is now called Moser's trick.
\begin{acknowledgments}
This research was supported by the U.S. Department of Energy (DE-AC02-09CH11466)
and, in part, by DOE Grant No. DE-SC0016072 and ARPA-E Grant No. DE-AR0001554. 
\end{acknowledgments}

\bibliographystyle{apsrev4-2}
\bibliography{Refs}

\end{document}